\documentclass[conference]{IEEEtran}
\IEEEoverridecommandlockouts
\usepackage{cite}
\usepackage{comment}
\usepackage{amsmath,amssymb,amsfonts}
\usepackage{algorithmic}
\usepackage{graphicx}
\usepackage{textcomp}
\usepackage{xcolor}
\usepackage{url}
\usepackage{balance}
\begin{document}

\title{Obstructed Target Tracking in Urban Environments}

\author{\IEEEauthorblockN{Christopher Berry, Donald J. Bucci Jr.}
\IEEEauthorblockA{\textit{Lockheed Martin Advanced Technology Labs}\\
Cherry Hill, NJ, USA \\
\texttt{christopher.m.berry@lmco.com}\\
\texttt{donald.j.bucci.jr@lmco.com}}
}

\maketitle

\begin{abstract}
Accurate tracking in urban environments necessitates target birth, survival, and detection models that quantify the impact of terrain and building geometry on the sequential estimation procedure.
Current efforts assume that target trajectories are limited to fixed paths, such as road networks.
In these settings a single airborne platform with a downward-facing camera is capable of fully observing a target, outside of a few obstructed regions that can be determined a priori (e.g. tunnels).
However, many practical target types are not necessarily restricted to road networks and thus require knowledge of azimuthal shadowed regions to the sensor.
In this paper, we propose the integration of geospatial data for an urban environment into a particle filter realization of a random finite set target tracking algorithm.
Specifically, we use 3D building polygons to compute the azimuthal shadowed regions with respect to deployed sensor location.
The particle filter predict and update steps are modified such that (1) target births are assumed to occur in line-of-sight (LOS) regions, (2) targets do not move into obstructions, (3) true target detections only occur in LOS regions.
The localization error performance improvement for a single target Bernoulli filter under these modifications is presented using freely available building vector data of New York City.
\end{abstract}

\section{Introduction}\label{sec:introduction}
Multi-target tracking in the context of urban surveillance applications is an increasingly important research area with many practical applications.
For example, closed-circuit television (CCTV) networks have been proposed for tracking objects as they move throughout a city \cite{Chen2015,Anuj2017}.
These objects can include persons-of-interest or vehicles.
The impact of target occlusion by buildings and other obstacles in the urban environment has a major impact on tracking performance when using these types of electro-optical (EO) sensors.
That is, as targets move behind obstacles the corresponding tracks are deprived of measurement updates.
This lack of measurements leads to progressively lower probability of existence and larger uncertainty volumes for the corresponding track estimates.
The estimated track is discarded if the target remains occluded for long enough, necessitating re-acquisition once it re-enters a line-of-sight (LOS) region.

A natural solution to this problem is to determine when the target is entering a non-line-of-sight (NLOS) region and modify the tracker update step to persist the corresponding track estimates for longer.
This problem of \emph{occlusion-aware tracking} has been investigated extensively in the computer vision community, where the goal is to detect a target and track its line of bearing \cite{Zhanli2015,Yang2015,Xu2015}.
The approach in \cite{Shin2017} describes a multi-sensor track management approach for pedestrian tracking on a vehicle mounted system.
Other related applications consist of overhead sensing schemes for persisting target tracks under partial occlusion within a priori available maps of road networks \cite{Ulmke2006,Ulmke2010}.
The authors specifically in \cite{Ulmke2010} assume that road segments are modeled using mixtures of Gaussians, and the shadowed segments of road are modeled by changing probability of detection for the corresponding subcomponents.
A similar approach was taken to track a single target on road networks in \cite{Vo2012}, but the authors leveraged the Bernoulli filter to model single target probability of existence.
Finally, recent work in \cite{Yi2013,Yi2010} suggested a data adaptive technique for estimating LOS/NLOS regions by associating the observed measurement error with a specified model for the LOS/NLOS areas (e.g., NLOS regions generating higher measurement error).
These techniques produced a probability estimate as to whether or not the measurement was generated from a LOS region.
Only measurements classified as coming from LOS regions were used to birth and update target tracks.

In this paper, we provide a method for integrating urban geographic information system (GIS) data into a Bernoulli random finite set (RFS) tracker \cite{Mahler2014} for localizing targets within a 2D cartesian plane.
Specifically, we use azimuthal range-bearing measurements as obtained from a notional stereoscopic EO sensor.
A particle filter implementation of the single target Bernoulli filter \cite{Ristic2013} is applied, where the target can either be detected or miss detected within uniformly distributed measurement clutter.
The methods discussed here are primarily focused on the particle filter importance sampling and detection probability estimation procedures.
Therefore, extension to the multi-target case under labeled RFS (e.g., $\delta$-GLMB \cite{Vo2014} LMB \cite{Reuter2014}) is straight forward.
As opposed to prior approaches for NLOS-enabled track-management schemes \cite{Shin2017}, the RFS framework is very attractive as it provides a unifying framework for modeling sensor measurement error, probability of detection, probability of survival, expected kinematics, and target birth distributions.
Although \cite{Ulmke2010} has also proposed the use of map-assisted tracking under occlusion, the Bernoulli and labeled RFS tracking algorithms provide direct estimates for probability of target existence.
These probability of existence estimates clearly indicate how the resulting track is maintained even when it enters NLOS regions.

The remainder of the paper is organized as follows.
In Section~\ref{sec:background}, we provide a brief overview of the RFS formalism and the particle filter implementation of the single target Bernoulli filter.
In Section~\ref{sec:terrain}, we discuss the geospatial data models considered by this work and how the predict and update steps of the Bernoulli particle filter are modified to incorporate this information.
Simulation results are provided in Section~\ref{sec:simulation} using open access building shapefile data of New York City \cite{nycmetadata}.
Specifically, we compare the tracker estimated probability of existence and the optimal subpattern assignment (OSPA) metrics achieved with and without the use of geospatial data.
The results indicate that target tracks are forward propagated in NLOS regions without significant degradation of localization accuracy or loss of track.

\section{Background}\label{sec:background}
A full treatment of the RFS formalism is given in \cite{Mahler2014}, and beyond the scope of this paper.
We follow the notation provided by \cite[Chapter 2.5, 4.1]{Ristic2013Book} that describes the particle filter implementation of the standard Bernoulli filter.

\subsection{Single Target Tracking with Random Finite Sets}
The multi-target tracking problem uses a set of measurements $\mathbf{Z}_k = \{z_{k,1},\ldots,z_{k,m_k}\} \in \mathcal{F}(\mathcal{Z})$ obtained at a discrete time step $k = 0,1,\ldots$ to estimate the state of a time varying number of targets, denoted $\mathbf{X}_k = \{x_{k,1},\ldots,x_{k,n_k}\}\in \mathcal{F}(\mathcal{X})$.
Here, the spaces $\mathcal{X}$ and $\mathcal{Z}$ denote the spaces of target states and measurements, and $\mathcal{F}(\cdot)$ represents the collection of finite subsets on a space.
The number of targets at each time step, $n_k$ are unknown and observed partially through the measurements.
The number of observed measurements at each time step is denoted $m_k$.
The measurement detection process is not ideal, implying in general that $m_k \neq n_k$.
For example, measurements generated by targets may be miss detected, resulting in $m_k < n_k$.
Measurements may also be the result of uncorrelated clutter, resulting in $m_k > n_k$.
For the multi-target case, the association between measurements and targets at each time step is also unknown.

The RFS formalism combines traditional probability theory along with concepts from the point process literature to address this problem in a common framework.
Although there are numerous axiomatic definitions \cite{Mahler2014}, we adopt the terminology that the set $\mathbf{X}$ is a RFS if its cardinality, $|\mathbf{X}|$, in addition to its individual elements are random variables.
This implies that a RFS is completely described by a cardinality distribution $P(|\mathbf{X}| = n_k)$ and a family of joint probability distributions, $f_{n_k}(\mathbf{x}_1,\ldots,\mathbf{x}_{n_k})$.
Applied more specifically to the multi-target tracking problem, the cardinality distribution associated with targets is largely determined by pre-defined by birth, survival, detection, and clutter stochastic processes.
The families of joint probability distributions are defined by the uncertainty models of the target kinematics and the sensor measurement likelihood functions.

A \emph{Bernoulli RFS} describes a special case where $|\mathbf{X}_k| \leq 1$ (i.e., at most one target). In this case, the RFS distribution function is
\begin{equation}
f(\mathbf{X}_k) = \begin{cases}
1-q & \mathbf{X}_k = \emptyset \\
qf(\mathbf{x}_k) & \mathbf{X}_k = \{\mathbf{x}_k\} \\
0 & \textnormal{otherwise}
\end{cases}
\end{equation}
where $q \in [0,1]$ is the probability of target existence and $f(\mathbf{x}_k)$ is the joint state distribution for a single target at time step $k$.
The multi-target case is constructed by taking the union of a set of Bernoulli RFS, and is denoted as a multi-Beroulli RFS \cite[Chapter 2.4.1]{Ristic2013Book}.
The multi-Bernoulli RFS, combined with a labeling scheme, is the building block used in the current state-of-the-art multi-target tracking filters using the RFS framework \cite{Vo2014}.

\subsection{Bernoulli Filter}
The multi-target Bayes filter in the RFS formalism is computationally intractable to implement for most cases.
For the case of a single target, however, the exact multi-target Bayes filter is the Bernoulli filter.
This filter captures a single object dynamic system that exhibits switching behavior in light of uncorrelated measurement observations due to clutter.
A full treatment of the RFS multi-target Bayes filter and the derivation of the Bernoulli filter is provided in \cite[Chapter 2.5]{Ristic2013Book} and \cite{Mahler2014}.
We present only the relevant prediction and update equations here for brevity.

The two posterior quantities of interest in the Bernoulli filter at time step $k$ are the probability of existence, denoted $q_{k|k}$, and the single target spatial distribution, denoted $s_{k|k}(\mathbf{x})$.
Let $p_b,p_s \in [0,1]$ be the probabilities that an object is born or survives at each time step.
Additionally, let $b_{k|k-1}(\mathbf{x})$ denote the spatial distribution of target births and $\pi_{k|k-1}(\mathbf{x}|\mathbf{x}')$ denote the spatial state dynamics (i.e., kinematic uncertainty model) for surviving targets.
At time step $k-1$, the \emph{predict step} is performed by propagating the probability of existence and spatial target distributions to time step $k$ via
\begin{equation}\label{eq:pred_q}
q_{k|k-1} = p_b(1-q_{k-1|k-1}) + p_sq_{k-1|k-1}
\end{equation}
\begin{multline}\label{eq:pred_s}
s_{k|k-1}(\mathbf{x}) = \frac{1}{q_{k|k-1}}\biggl(p_b(1-q_{k-1|k-1})b_{k|k-1}(\mathbf{x})\\
+p_sq_{k-1|k-1}\int \pi_{k|k-1}(\mathbf{x}|\mathbf{x}')s_{k-1|k-1}(\mathbf{x}')d\mathbf{x}' \biggr).
\end{multline}

Let $p_D(\mathbf{x})$ represent the probability of detection model, and $g(\mathbf{x}|\mathbf{z})$ the sensor measurement likelihood function.
An additional assumption is added such that the measurement clutter process is represented as a Poisson RFS having arrival intensity $\lambda$ and spatial distribution (in the measurement space) $c(\mathbf{z})$.
The \emph{update step} is then executed by applying the RFS analogue of Bayes' rule, resulting in
\begin{equation}\label{eq:upd_q}
q_{k|k} = \frac{1-\Delta_k}{1-\Delta_kq_{k|k-1}}q_{k|k-1}
\end{equation}
\vspace*{0.005in}
\begin{multline}\label{eq:upd_s}
s_{k|k}(\mathbf{x}) = \frac{1}{1-\Delta_k}\Biggl(1-p_D(\mathbf{x})\\
+ p_D(\mathbf{x})\sum_{\mathbf{z} \in \mathbf{Z}_k} \frac{g(\mathbf{z}|\mathbf{x})}{\lambda c(\mathbf{z})}\Biggr)s_{k|k-1}(\mathbf{x}),
\end{multline}
where
\begin{multline}\label{eq:delta}
\Delta_k = \int_\mathcal{X} p_D(\mathbf{x})s_{k|k-1}(\mathbf{x})d\mathbf{x} \\ - \sum_{\mathbf{z} \in \mathbf{Z}_k} \int_\mathcal{X} p_D(\mathbf{x}) \frac{g(\mathbf{z}|\mathbf{x})}{\lambda c(\mathbf{z})}s_{k|k-1}(\mathbf{x})d\mathbf{x}.
\end{multline}

The Bernoulli filter predict and update steps have an intuitive explanation.
The predicted probability of existence is a weighted combination of two events: the target survives from the previous time step or the target from the previous time step dies and is reborn at the next time step.
The spatial target distribution is the weighted combination of the birth distribution (if the target dies) and the standard Chapman-Kolmogorov prediction integral \cite{Sheldon2014} based on the kinematic model (if the target survives).
The update step involves a weighted combination of the events that the target is not detected, or the target is detected and assigned each of the observed measurements.
Because of the single target assumption, a measurement that is not assigned to the target must be the result of clutter.

\subsection{Sequential Monte Carlo Implementation}
Implementing the Bernoulli filter predict and update steps requires evaluating the corresponding integrals in Equations~(\ref{eq:pred_s})-(\ref{eq:delta}).
In order to flexibly incorporate the geospatial data models necessitated by this paper, we leverage the particle filter implementation presented in \cite[Chapter 4.1]{Ristic2013Book}.
Specifically, the spatial distribution at time step $k-1$ is approximated using a sufficiently large particle system $\{(w_{k-1}^{(1)},\mathbf{x}_{k-1}^{(1)}),\ldots,(w_{k-1}^{(J)},\mathbf{x}_{k-1}^{(J)})\}$ such that
\begin{equation*}
s_{k-1|k-1}(\mathbf{x}) \approx \hat{s}_{k-1|k-1}(\mathbf{x}) = \sum_{i=1}^J w_{k-1}^{(i)}\delta_{\mathbf{x}_{k-1}^{(i)}}(\mathbf{x}),
\end{equation*}
where $\delta_\mathbf{a}(\mathbf{b}) = 1$ if $\mathbf{a} = \mathbf{b}$ and zero otherwise.
Given the probability of existence from the previous time step, Equation~(\ref{eq:pred_q}) can be applied directly.
The predict step for the spatial distribution of Equation~(\ref{eq:pred_s}) is executed using an importance sampling procedure,
\begin{equation}
\mathbf{x}_{k|k-1}^{(i)} \sim \begin{cases}
\rho(\mathbf{x}|\mathbf{x}_{k-1}^{(i)},\mathbf{Z}_k) & i = 1,\ldots,J \\
\beta(\mathbf{x}|\mathbf{Z}_k) & i = J+1,\ldots,J+B
\end{cases},
\end{equation}
where $B$ is the number of birth particles generated.
The functions $\rho(\cdot)$ and $\beta(\cdot)$ define proposal distributions for the survival and birth process that are designed to be easily sampled from.
The corresponding weights are calculated as
\begin{equation}
\mathbf{w}_{k|k-1}^{(i)} = \frac{p_sq_{k-1|k-1}}{q_{k|k-1}}\frac{\pi_{k|k-1}(\mathbf{x}_{k|k-1}^{(i)}|\mathbf{x}_{k-1}^{(i)})}{\rho(\mathbf{x}_{k|k-1}|\mathbf{x}_{k-1}^{(i)},\mathbf{Z}_k)}w_{k-1}^{(i)}
\label{eq:survival_weight}
\end{equation}
for $i=1,\ldots,J$ and
\begin{equation}
\mathbf{w}_{k|k-1}^{(i)} = \frac{p_b(1-q_{k-1|k-1})}{q_{k|k-1}}\frac{b_{k|k-1}(\mathbf{x}_{k|k-1}^{(i)})}{\beta(\mathbf{x}_{k|k-1}^{(i)}|\mathbf{Z}_k)}\frac{1}{B}
\label{eq:birth_weight}
\end{equation}
for $i=J+1,\ldots,J+B$. The update step is performed by first approximating the normalization constant $\Delta_k$ as
\begin{multline}
\hat{\Delta}_k = \sum_{i=1}^{J+B} p_D(\mathbf{x}_{k|k-1}^{(i)})\mathbf{w}_{k|k-1}^{(i)} \\ -  \sum_{\mathbf{z} \in \mathbf{Z}_k}\sum_{i=1}^{J+B} p_D(\mathbf{x}_{k|k-1}^{(i)})\frac{g(\mathbf{z}|x_{k|k-1}^{(i)})}{\lambda c(\mathbf{z})}\mathbf{w}_{k|k-1}^{(i)}.
\end{multline}
The estimated value of $\hat{\Delta}_k$ is substituted in Equation~(\ref{eq:upd_q}) in place of $\Delta_k$ to get the posterior probability of existence.
The posterior spatial distribution weights are updated using Equation~(\ref{eq:upd_s}) as
\begin{multline}
w_{k|k}^{(i)} \propto \left(1-p_D(\mathbf{x}_{k|k-1}^{(i)}) \right. \\ \left.+ p_D(\mathbf{x}_{k|k-1}^{(i)})\sum_{\mathbf{z} \in \mathbf{Z}_k}\frac{g(\mathbf{z}|x_{k|k-1}^{(i)})}{\lambda c(\mathbf{z})}\right)\mathbf{w}_{k|k-1}^{(i)}
\end{multline}
such that the resulting values are normalized to sum to one.
Finally, a resampling procedure is applied to generate a new size $J$ particle system from $\{(w_{k|k}^{(1)},\mathbf{x}_{k|k-1}^{(1)}),\ldots,(w_{k|k}^{(J+B)},\mathbf{x}_{k|k-1}^{(J+B)})\}$.
For more detailed implementation notes, we refer the reader to the pseudocode available in \cite[Chapter 4.1, Algorithm 3]{Ristic2013Book} and in \cite{Ristic2013}.

\section{Geospatial-dependent Filtering Method}\label{sec:terrain}

\subsection{Geospatial Model}\label{sec:shapefile}
\begin{figure}
\centering
\includegraphics[width=0.48\textwidth]{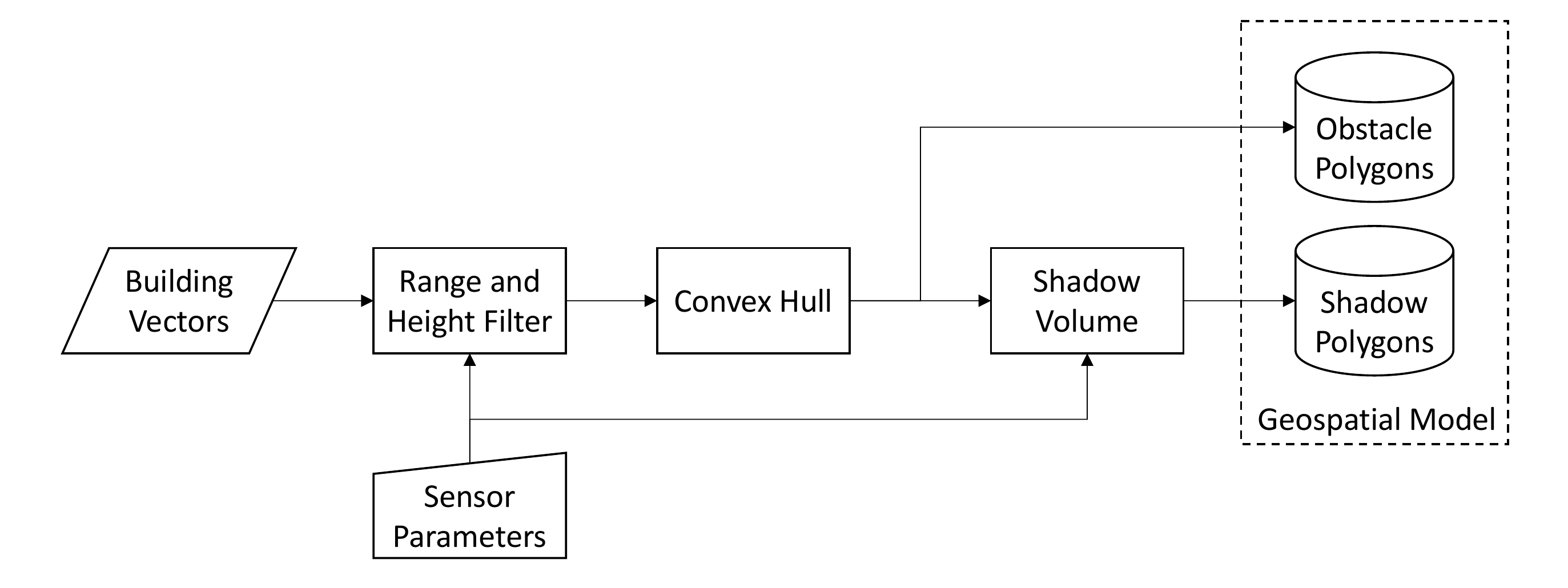}
\caption{Block diagram for creating geospatial model.}
\label{fig:block_diagram}
\end{figure}

A block diagram of the offline process used to create the geospatial model from GIS data is shown in Figure~\ref{fig:block_diagram}.
GIS data is publicly available from government sources, typically as raster data of elevation from aerial imagery or LIDAR scans.
Advances in sensing and image processing has enabled photogrammetric extraction high accuracy building footprints even in dense urban canyons.
For the work in this paper, we use building footprints from \cite{nycmetadata}, available in shapefile format.
Shapefiles are desirable as each building perimeter outline is represented as geospatial vector data, which allows computationally efficient geometric calculations.
Each building vector also has associated metadata, including ground and roof heights.

Buildings located beyond the range and below the height of the sensor are filtered out.
We assume a building can be approximated by a convex shape without negligible loss of accuracy.
To ensure this, each building is transformed into into a new 2D polygon via its convex hull.
A convex hull is the smallest convex set containing a set of points.
The convex hull operation is implemented in Python using the Shapely package.
The convex building polygons are stored as the set of obstacles that particles may not persist in.

A shadow volume algorithm is then used to project hard shadows cast by buildings from the sensor in a 2D plane.
A shadow volume is constructed by extending rays directed from the sensor position through each convex building vertex to the edge of the surveillance region.
The shadow polygons formed from each building are stored as the set of NLOS regions that particles are miss detected in.

\subsection{Filtering Method}
We now describe how to incorporate the geospatial model into the Bernoulli filter.
Knowledge of building structures allows the filter to constrain predicted target positions to areas that are not contained within a building.
Kinematic models typically used in target tracking, such as constant velocity or constant turn \eqref{eq:constant_turn}, do not place any restrictions on the state space.
Instead, the transition density in Equation~\eqref{eq:survival_weight} is modified to apply zero weight to positions inside buildings as
\begin{equation}
\pi_{k|k-1}(\mathbf{x}_{k|k-1}^{(i)}|\mathbf{x}_{k-1}^{(i)}) =
\begin{cases}
\begin{matrix}
\epsilon & \mathbf{x}_{k|k-1}^{(i)} \text{in obstacles}\\
\pi_{k|k-1}(\cdot) & \text{otherwise}\\
\end{matrix}
\end{cases}
\end{equation}
where $\epsilon$ is very small.
It is important to note that the geospatial model is incorporated into the survival model to reduce uncertainty and not to model target death by lowering the sum total of all predicted weights.
Therefore, it is necessary to normalize the weights by $(p_S q_{k-1|k-1}/(q_{k|k-1}\Sigma w_k^{(i)}))^{-1}$ for $i=1,\ldots,J$, or resample $\mathbf{x}_{k|k-1}^{(i)}$ for all $\mathbf{w}_{k|k-1}^{(i)} = \epsilon$ if sample impoverishment is severe.

The sensor is not able to detect the target when it is occluded by buildings and terrain.
The obstructed regions calculated offline in Section~\ref{sec:shapefile} can be used by the filter in evaluation of the state-dependent probability of detection.
Intuitively, the detection model should be zero for particles in occluded regions.
That is,
\begin{equation}
p_D(\mathbf{x}_{k|k-1}^{(i)}) =
\begin{cases}
\begin{matrix}
\epsilon & \mathbf{x}_{k|k-1}^{(i)} \text{in shadow polygon}\\
p_D(\cdot) & \text{otherwise}\\
\end{matrix}
\end{cases}.
\label{eq:pD_NLOS}
\end{equation}
By conditioning $p_D(\mathbf{x}_{k|k-1}^{(i)})$ on whether $\mathbf{x}_{k|k-1}^{(i)}$ is occluded, the filter is able to persist the estimated track through successive missed detections, which would have otherwise reduced the probability of existence to zero.

The spatial distribution of target births must also be modified to incorporate the geospatial data model.
The proposal distribution $\beta(\mathbf{x}_{k|k-1}^{(i)}|\mathbf{Z}_k)$ may be static, based on prior information or uniform if no information is available, or adaptive, based on the measurements.
The birth density must be consistent with the sensor model, which implies that targets are born in LOS regions to the target.
If the target is born in NLOS regions, it is undetectable by the sensor until it is in a LOS region.
The birth density is modified such that $\mathbf{w}_{k|k-1}^{(i)}$ for $i=J+1,\ldots,J+B$ are zero in occluded positions as
\begin{equation}
b_{k|k-1}(\mathbf{x}_{k|k-1}^{(i)}) =
\begin{cases}
\begin{matrix}
\epsilon & \mathbf{x}_{k|k-1}^{(i)} \text{in shadow polygon}\\
b_{k|k-1}(\cdot) & \text{otherwise}\\
\end{matrix}
\end{cases}.
\end{equation}
Similarly to the survival process, the birth weights $\mathbf{w}_{k|k-1}^{(i)}$ for $i=J+1,\ldots,J+B$ are normalized by $(p_b(1-q_{k-1|k-1})/(q_{k|k-1}\Sigma w_k^{(i)}))^{-1}$, or resampled $\mathbf{x}_{k|k-1}^{(i)}$ for all $\mathbf{w}_{k|k-1}^{(i)} = \epsilon$ if sample impoverishment is severe.
The rational for making the birth density state-dependent is that it prevents selection of particles born in NLOS regions from persisting, which would not be filtered out in the update step.
Otherwise the probability of existence would be raised artificially and initial state estimates when the target is acquired would be biased.

\section{Tracking Simulation}\label{sec:simulation}
\subsection{Simulation Description}\label{sec:setup}
\begin{figure}
\centering
\includegraphics[width=0.48\textwidth]{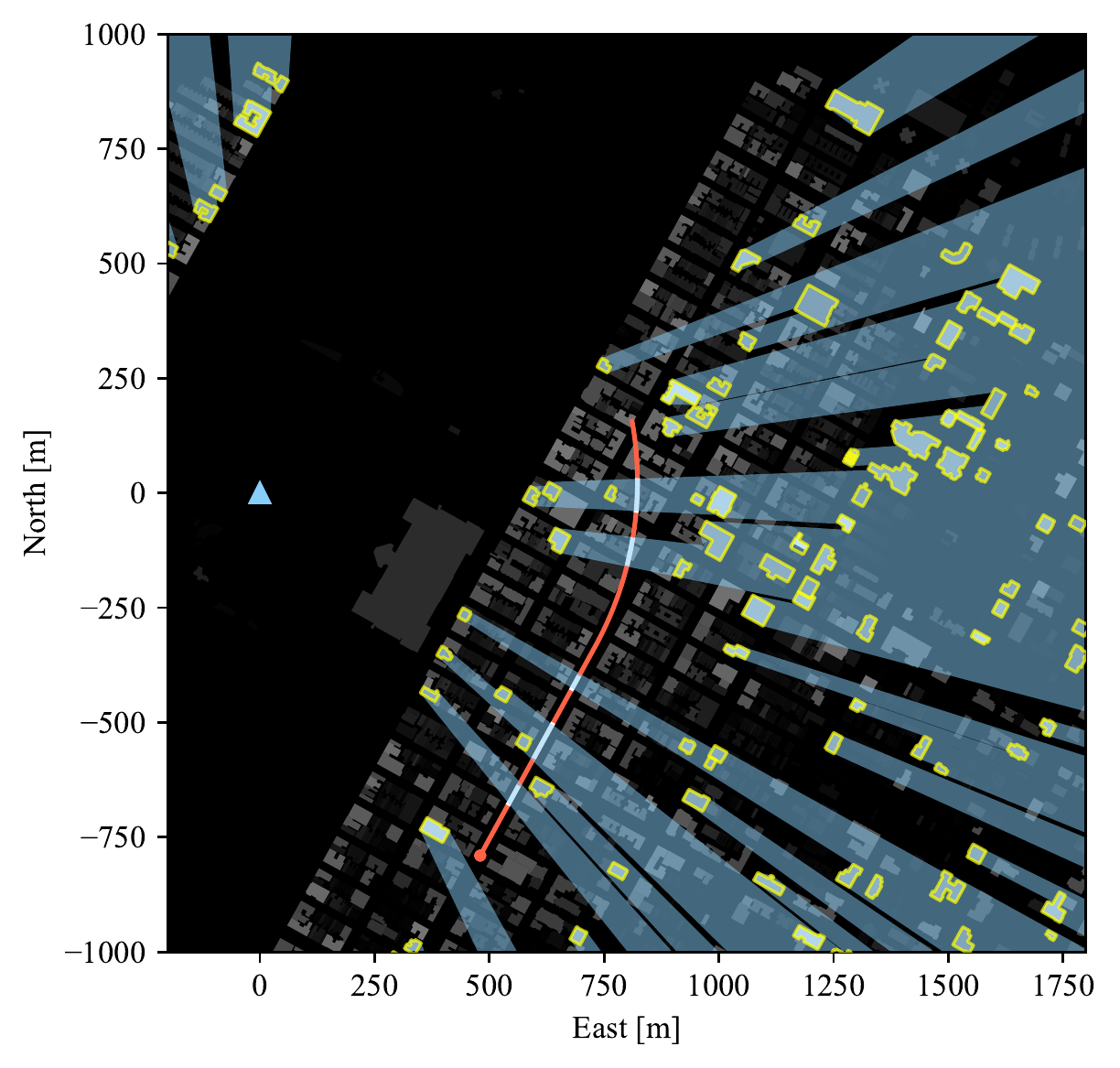}
\caption{Simulation in 2D ENU plane.
Sensor position denoted by $\blacktriangle$.
Buildings (\textit{grey}) which occlude the sensor's FOV are outlined (\textit{yellow}) with projection of shadows (\textit{blue}).
Target trajectory segmented by LOS (\textit{red}) and NLOS (\textit{white}); starting point denoted by $\bullet$.}
\label{fig:laydown}
\end{figure}

In this section we numerically compare the tracking performance of the Bernoulli filter with and without the geospatial model.
Figure~\ref{fig:laydown} shows the simulation scenario.
Building vertices were converted from geodetic coordinates to East, North, Up (ENU) using a reference point of ($-73.9675^\circ, 40.781^\circ, 200$m).
The stationary sensor was positioned at the origin in Central Park.
Buildings with heights above $115$m were incorporated in the geospatial model.
The simulation did not assume any error between the geospatial model and the actual locations of the buildings.

The target travelled along a street in Manhattan at $13$m/s, in the same elevation plane as the sensor.
Its birth and death occurred in the sensor's LOS region.
The target state vector consisted of $x_k = [y_k^T, \omega_k]$, where $y_k = [p_\text{East}, \dot{p}_\text{East}, p_\text{North}, \dot{p}_\text{North}]$ and $\omega_k$ is the turn rate.
Target motion followed a constant turn model, given by
\begin{equation}
\begin{matrix}
y_k =           & F(\omega_{k-1})y_{k-1} + G w_{k-1}\\
\omega_k = & \omega_{k-1} + T w_{k-1}\\
\end{matrix}
\label{eq:constant_turn}
\end{equation}
where
$$
F(\omega) =  \begin{bmatrix}
1 &  \frac{\sin\omega T}{\omega}      & 0 & -\frac{1 - \cos\omega T}{\omega}\\
0 &  \cos\omega T                             & 0 & -\sin\omega T\\
0 & \frac{1 - \cos\omega T}{\omega} & 1 & \frac{\sin\omega T}{\omega}\\
0 &  \sin\omega T                              & 0 & \cos\omega T\\
\end{bmatrix},
G = \begin{bmatrix}
\frac{T^2}{2} & 0\\
T                  & 0\\
0                  & \frac{T^2}{2}\\
0                  & T\\
\end{bmatrix}
$$
The simulation used a sampling interval of $T = 1$s, process noise $w_k \sim \mathcal{N}(\cdot; 0, \sigma_w^2 I_2)$, $\sigma_w = 2.5 \text{ m/s}^2$, $I_n$ is the $n \times n$ identity matrix, and $u_k \sim \mathcal{N}(\cdot; 0, \sigma_u^2)$,  $\sigma_u = \pi/180 \text{ rad/s}$.
At time step $k=46$, the target initiated a $\omega=\pi/180$ rad/s turn.

\begin{figure}
\centering
\includegraphics[width=0.48\textwidth]{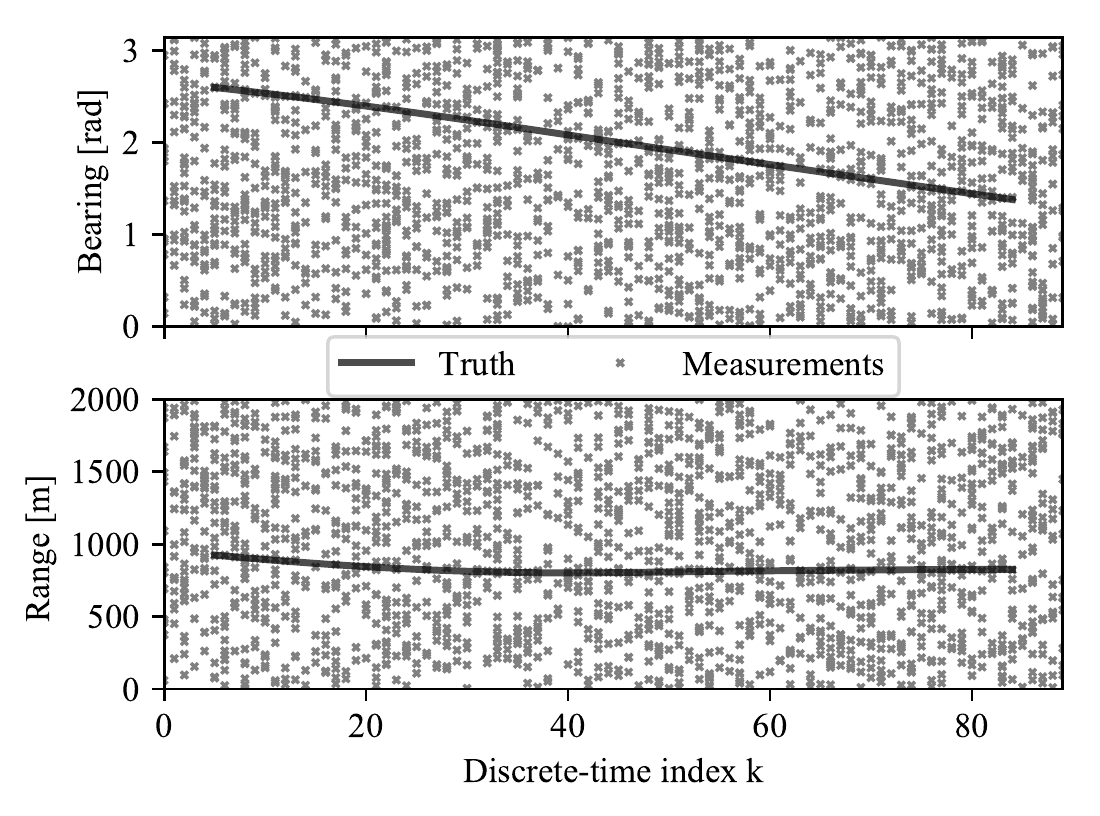}
\caption{Single realization of observed range-bearing measurements ($\times$) over time.
Target trajectory in measurement space shown as solid line.}
\label{fig:cluttermeas}
\end{figure}

A generic bearing-range sensor returned observations
\begin{equation}
z_k = \begin{bmatrix}
\text{arctan}\left(\frac{p_\text{East}}{p_\text{North}}\right)\\
\sqrt{p_\text{East}^2 + p_\text{North}^2}
\end{bmatrix} + v_k
\label{eq:bearing_range}
\end{equation}
where $v_k \sim \mathcal{N}(\cdot; 0, R_k)$ with $R_k = \text{diag}\left(\begin{bmatrix} \sigma_\theta^2, \sigma_r^2 \end{bmatrix}\right)$, $\sigma_\theta = 2\pi/180$ rad/s and $\sigma_r = 10$ m.
The target was detected with probability of detection $p_D = 0.98$ when the path from the sensor to the target was unobstructed.
Clutter was uniformly distributed over the area of interest $\begin{bmatrix} 0, \pi \end{bmatrix}$ rad $\times \begin{bmatrix} 0, 2000 \end{bmatrix}$ m (\textit{i.e.}, $c(z) = (2000\pi)^{-1}$), with the number of clutter returns Poisson distributed with $\lambda = 20$.
Figure~\ref{fig:cluttermeas} shows the measurements over time.

The Bernoulli particle filter used $N = 5000$ particles.
The probability of target survival and birth were set to $p_S = 0.98$ and $p_B = 0.01$, respectively.
Particles were adaptively birthed using the technique described in \cite{Ristic2013}, with $B = N$, $\sigma_v = 10$ m/s, and $\sigma_\omega = 30\pi/180$ rad.

\subsection{Monte Carlo Results}
\begin{figure}
\centering
\includegraphics[width=0.48\textwidth]{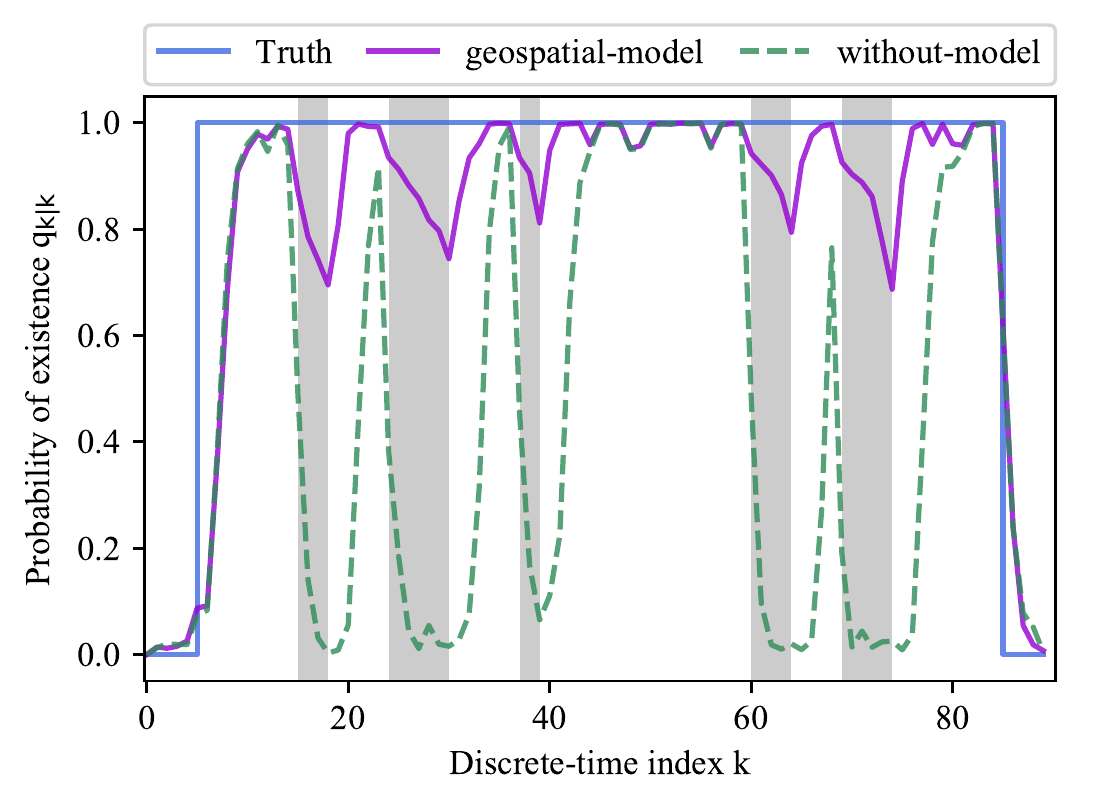}
\caption{True cardinality versus Bernoulli filter probability of existence averaged over 100 MC iterations.
Shaded segments denote time steps in which the target is obstructed by buildings.}
\label{fig:card}
\end{figure}

The performance of the Bernoulli filter with and without the geospatial model was evaluated over $100$ Monte Carlo (MC) simulations.
Figure~\ref{fig:card} shows the probability of existence at each time averaged over all MC iterations of both filters against ground truth.
The five shadowed regions the target travels through, highlighted in grey blocks, correspond to the dips in $q_{k|k}$ of both filters starting at $k = $15, 24, 37, 60, and 69.
However, the filter with the geospatial model is robust to missed detections and maintains a high probability of existence.
The filter without the geospatial model completely loses track of the target and suffers a delay reacquiring the target each time it is detectable.

While the geospatial model should allow the Bernoulli filter to maintain high probability of existence during segments of the target trajectory in shadowed regions, two effects are responsible for $q_{k|k} < 1$.
The first decrease of $q_{k|k}$ occur at the boundary of the detectable region and is explained as follows:
during the obstructed segments, the spatial distribution of the particle system spreads out with each successive predict step and effectively no update as $p_D \approx 0$.
As particles cross into the detectable region before the target, the missing measurements reduce particle weight by $1 - p_D$, which lowers the overall probability of existence update.
Similarly, particles lagging behind the true target position as it becomes shadowed will be weighted $1 - p_D$.

The second decrease of $q_{k|k}$ is due to non-unity probability of survival, the effect of which compounds the longer the target is in a NLOS region.
One solution would be to set $p_S = 1$ when the target is not detectable.
However, this would have the undesirable effect of artificially raising $q_{k|k}$ as particles diverge in NLOS regions.
This can be seen in Figure~\ref{fig:card} during the last 10 times steps of the filter with geospatial model, where particles have moved behind the next building and  $q_{k|k}$ is nonzero.

\begin{figure}
\centering
\includegraphics[width=0.48\textwidth]{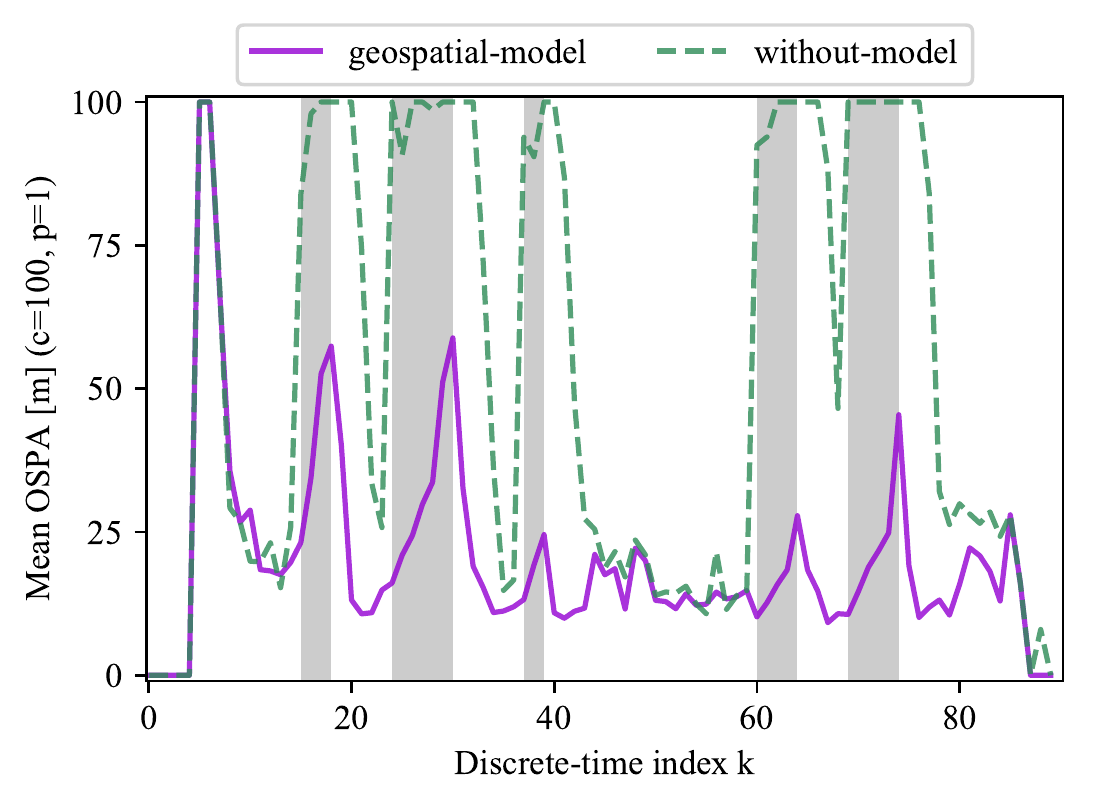}
\caption{OSPA metric over time averaged over 100 MC iterations for Bernoulli filter with (\textit{purple}) and without (\textit{dashed green}) geospatial model.
Shaded segments denote time steps in which the target is obstructed by buildings.}
\label{fig:ospa}
\end{figure}

The optimal subpattern assignment (OSPA) distance \cite{Schuhmacher2008b} for both filters is shown Figure~\ref{fig:ospa}.
The cutoff parameter was set to $c = 100\text{m}$.
The dominating source of error for the filter without the geospatial model was cardinality.
When the target was detectable and both filters were reporting estimates on it, their localization error was comparable.

\section{Conclusion}
In this paper we presented a method for incorporating geospatial data of building heights and locations into a hypothetical stereoscopic EO (i.e., range-bearing), single target tracking system.
Specifically, we suggested techniques for using the NLOS and obstructed map regions of an urban environment to prevent the birth and survival importance sampling procedures from generating particles within buildings or shadowed regions.
In the update step, we proposed setting particle weights using a negligible probability of detection.
Simulation results demonstrated that this shadow-aware technique is capable of persisting targets for longer durations of time while they remain shadowed.
As a result, the system does not need to perform track re-acquisition as targets enter and leave NLOS regions.

\balance
\bibliographystyle{IEEEtran}
\bibliography{SSL-2019-Fusion_Shadow.bib}

\begin{thebibliography}{10}
\providecommand{\url}[1]{#1}
\csname url@samestyle\endcsname
\providecommand{\newblock}{\relax}
\providecommand{\bibinfo}[2]{#2}
\providecommand{\BIBentrySTDinterwordspacing}{\spaceskip=0pt\relax}
\providecommand{\BIBentryALTinterwordstretchfactor}{4}
\providecommand{\BIBentryALTinterwordspacing}{\spaceskip=\fontdimen2\font plus
\BIBentryALTinterwordstretchfactor\fontdimen3\font minus
  \fontdimen4\font\relax}
\providecommand{\BIBforeignlanguage}[2]{{%
\expandafter\ifx\csname l@#1\endcsname\relax
\typeout{** WARNING: IEEEtran.bst: No hyphenation pattern has been}%
\typeout{** loaded for the language `#1'. Using the pattern for}%
\typeout{** the default language instead.}%
\else
\language=\csname l@#1\endcsname
\fi
#2}}
\providecommand{\BIBdecl}{\relax}
\BIBdecl

\bibitem{Chen2015}
Z.~Chen, W.~Liao, B.~Xu, H.~Liu, Q.~Li, H.~Li, C.~Xiao, H.~Zhang, Y.~Li,
  W.~Bao, and D.~Yang, ``Object tracking over a multiple-camera network,'' in
  \emph{IEEE International Conference on Multimedia Big Data}, 2015.

\bibitem{Anuj2017}
L.~Anuj and M.~T.~G. Krishna, ``Multiple camera based multiple object tracking
  under occlusion: {A} survey,'' in \emph{International Conference on
  Innovative Mechanisms for Industry Applications (ICIMIA)}, 2017.

\bibitem{Zhanli2015}
L.~{Zhanli}, C.~{Leilei}, and X.~{Ailing}, ``Target tracking algorithm based on
  particle filter and mean shift under occlusions,'' in \emph{IEEE
  International Conference on Signal Processing, Communications and Computing
  (ICSPCC)}, Sep. 2015, pp. 1--4.

\bibitem{Yang2015}
B.~Yang and R.~Yang, ``Interactive particle filter with occlusion handling for
  multi-target tracking,'' in \emph{12th International Conference on Fuzzy
  Systems and Knowledge Discovery (FSKD)}, Aug 2015, pp. 1945--1949.

\bibitem{Xu2015}
Y.~{Xu}, X.~{Jiang}, and F.~{Li}, ``Improved anti-occlusion target tracking
  algorithm based on compressive particle filtering,'' in \emph{8th
  International Symposium on Computational Intelligence and Design (ISCID)},
  Dec 2015, pp. 527--531.

\bibitem{Shin2017}
S.~{Shin}, D.~{Ahn}, and H.~{Lee}, ``Occlusion handling and track management
  method of high-level sensor fusion for robust pedestrian tracking,'' in
  \emph{IEEE International Conference on Multisensor Fusion and Integration for
  Intelligent Systems (MFI)}, Nov 2017, pp. 233--238.

\bibitem{Ulmke2006}
M.~{Ulmke} and W.~{Koch}, ``Road-map assisted ground moving target tracking,''
  \emph{IEEE Transactions on Aerospace and Electronic Systems}, vol.~42, no.~4,
  pp. 1264--1274, October 2006.

\bibitem{Ulmke2010}
M.~{Ulmke}, O.~{Erdinc}, and P.~{Willett}, ``{GMTI} tracking via the {G}aussian
  mixture cardinalized probability hypothesis density filter,'' \emph{IEEE
  Transactions on Aerospace and Electronic Systems}, vol.~46, no.~4, pp.
  1821--1833, October 2010.

\bibitem{Vo2012}
B.~T. {Vo}, C.~M. {See}, N.~{Ma}, and W.~T. {Ng}, ``Multi-sensor joint
  detection and tracking with the {B}ernoulli filter,'' \emph{IEEE Transactions
  on Aerospace and Electronic Systems}, vol.~48, no.~2, pp. 1385--1402, April
  2012.

\bibitem{Yi2013}
L.~{Yi}, S.~G. {Razul}, Z.~{Lin}, and C.~{See}, ``Individual {AOA} measurement
  detection algorithm for target tracking in mixed {LOS/NLOS} environments,''
  in \emph{IEEE International Conference on Acoustics, Speech and Signal
  Processing}, May 2013, pp. 3924--3928.

\bibitem{Yi2010}
L.~{Yi}, C.~{Lim}, C.~{See}, S.~G. {Razul}, and Z.~{Lin}, ``Robust tracking in
  mixed {LOS/NLOS} environments,'' in \emph{11th International Conference on
  Control Automation Robotics Vision}, Dec 2010, pp. 497--500.

\bibitem{Mahler2014}
R.~Mahler, \emph{Advances in Statistical Multisource-Multitarget Information
  Fusion}.\hskip 1em plus 0.5em minus 0.4em\relax Artech House, 2014.

\bibitem{Ristic2013}
B.~{Ristic}, B.~{Vo}, B.~{Vo}, and A.~{Farina}, ``A tutorial on {B}ernoulli
  filters: Theory, implementation and applications,'' \emph{IEEE Transactions
  on Signal Processing}, vol.~61, no.~13, pp. 3406--3430, July 2013.

\bibitem{Vo2014}
B.~{Vo}, B.~{Vo}, and D.~{Phung}, ``Labeled random finite sets and the {B}ayes
  multi-target tracking filter,'' \emph{IEEE Transactions on Signal
  Processing}, vol.~62, no.~24, pp. 6554--6567, Dec 2014.

\bibitem{Reuter2014}
S.~{Reuter}, B.~{Vo}, B.~{Vo}, and K.~{Dietmayer}, ``The labeled
  multi-{B}ernoulli filter,'' \emph{IEEE Transactions on Signal Processing},
  vol.~62, no.~12, pp. 3246--3260, June 2014.

\bibitem{nycmetadata}
\BIBentryALTinterwordspacing
{City of New York: NYC Geo Metadata}. [Online]. Available:
  \url{https://github.com/CityOfNewYork/nyc-geo-metadata/blob/master/Metadata/Metadata_BuildingFootprints.md}
\BIBentrySTDinterwordspacing

\bibitem{Ristic2013Book}
B.~Ristic, \emph{Particle Filters for Random Finite Set Models}.\hskip 1em plus
  0.5em minus 0.4em\relax Springer, 2013.

\bibitem{Sheldon2014}
S.~Ross, ``Chapter 4.2: {Chapman-Kolmogorov} equations,'' in \emph{Introduction
  to Probability Models}, 11st~ed.\hskip 1em plus 0.5em minus 0.4em\relax
  Academic Press, 2014.

\bibitem{Schuhmacher2008b}
D.~Schuhmacher, B.-T. Vo, and B.-N. Vo, ``A consistent metric for performance
  evaluation of multi-object filters,'' \emph{IEEE Transactions on Signal
  Processing}, vol.~56, no.~8, pp. 3447--3457, 2008.

\end{thebibliography}

\end{document}